\documentclass[aps,prl,reprint,superscriptaddress]{revtex4-1}
\usepackage{amssymb}
\usepackage{graphicx}
\usepackage{amssymb}
\usepackage{epsfig}
\usepackage{multirow}
\usepackage{color}
\usepackage{hyperref}
\usepackage{url}
\makeatletter
\g@addto@macro{\UrlBreaks}{\UrlOrds}
\g@addto@macro{\UrlBreaks}{%
\do\/\do\d%
}
\makeatother
\begin{document}
 
\title{Standard Model radiative corrections in the pion form factor measurements \\do not explain the $a_\mu$ anomaly}

\author{Francisco Campanario}
\affiliation{Instituto de F\'{\i}sica Corpuscular, CSIC-Universitat de Val\`encia,
 E-46980 Paterna, Valencia, Spain}
\author{Henryk Czy\.z}
\affiliation{Institute of Physics, University of Silesia, Katowice, Poland} 
%\affiliation{Institute of Physics, University of Silesia, PL-41500 Chorz\'ow, Poland}
\author{Janusz Gluza}
\affiliation{Institute of Physics, University of Silesia, Katowice, Poland}
\affiliation{Faculty of Science, University of Hradec Kr\'alov\'e, Czech Republic}
\author{Tomasz~Jeli\'nski}
\affiliation{Institute of Physics, University of Silesia, Katowice, Poland} 
\author{Germ\'an Rodrigo}
\affiliation{Instituto de F\'{\i}sica Corpuscular, CSIC-Universitat de Val\`encia,
E-46980 Paterna, Valencia, Spain}
\author{Szymon Tracz}
\affiliation{Institute of Physics, University of Silesia, Katowice, Poland} 
%\affiliation{Institute of Physics, University of Silesia, PL-41500 Chorz\'ow, Poland}
\author{Dmitry Zhuridov}
\affiliation{Institute of Physics, University of Silesia, Katowice, Poland} 
%\affiliation{Institute of Physics, University of Silesia, PL-41500 Chorz\'ow, Poland}
%\thanks{}


\date{\today}

\begin{abstract}
   In this letter, we address the question of whether the almost four standard deviations
  difference between theory and experiment for the muon anomalous magnetic moment $a_\mu$ can be explained as a higher-order Standard Model perturbation effect
  in the pion form factor measurements. This question has, until now,
  remained open, obscuring the source of discrepancies between the measurements.
  We calculate the last radiative corrections for the extraction of the pion form factor, which were believed to be potentially substantial enough to explain the data within the Standard Model.
We find that the corrections are too small to diminish existing discrepancies in the determination of the pion form factor for different kinematical configurations of low-energy  BABAR, BESS-III and KLOE experiments.
Consequently, they cannot noticeably change the previous predictions for $a_\mu$ and decrease the deviations between theory and direct measurements.
To solve the above issues, new data  and better understanding of low-energy experimental setups are needed, especially as new direct $a_\mu$ measurements at Fermilab and J-PARC will provide new insights and  substantially shrink the experimental error.
%Modelling the pion form factor constitutes the most relevant part of the hadronic contribution to the anomalous magnetic moment of the muon $a_\mu$, affecting strongly $a_\mu$ error estimate.
\end{abstract}

 
\maketitle
%\allowdisplaybreaks

\paragraph{Introduction.}
The anomalous magnetic moment of the muon $a_\mu \equiv (g_\mu-2)/2$ is predicted in the Standard Model (SM) with an accuracy at the level
of 0.3 ppm~\cite{Jegerlehner:2017gek,Davier:2017zfy,Keshavarzi:2018mgv,PhysRevD.98.030001} while the precision of the direct experimental measurement is of the order of 0.54 ppm~\cite{Bennett:2006fi}.  
Remarkably, the tension between the experimental measurement and the SM prediction, $a_\mu^{\rm exp} - a_\mu^{\rm SM} =  268(63)(43)  \times 10^{-11}$~\cite{PhysRevD.98.030001}, corresponds to about  3.5 standard deviations.
This is one of the largest and long-standing discrepancies between the SM and experiment. The central question is whether the discrepancy is due to unknown new physics effects beyond the SM (new particles and new interactions) or to theoretical and/or experimental errors not completely under control.
Concerning the beyond the SM option, there are a few models, which are able to shift the theoretical prediction for $a_\mu$ in the direction of the experimental value in selected regions of the parameter space~\cite{Bach:2016mhg,Cherchiglia:2017uwv}.  However, many commonly used non-standard models
have problems to accommodate this discrepancy and will have to be modified
or rejected,
 when that harbinger of new physics is confirmed~\cite{Czarnecki:2001pv}.  

In this letter, we scrutinize possible flaws in the estimation of the theoretical and experimental errors by re-investigating the SM input into data analysis related to pion-photon interactions, including so far missing and potentially relevant radiative corrections. %In this way we are aiming to solve inconsistencies among low-energy experiments, which impinge determination of $a_\mu$. 

This study is particularly timely due to the fact that new measurements at Fermilab~\cite{Lusiani:2018tcd} and J-PARC~\cite{Sato:2017sdn} 
aim to reduce the experimental error of the direct measurement by a factor of four. 
Therefore, the theoretical and experimental groups that contribute to the accurate determination of $a_\mu$ must point to a similar precision in the near future~\cite{g-2theory}, and also to understand definitely the source of the present discrepancies.

The QED and pure electroweak SM contributions to $a_\mu$ are known presently with a satisfactory precision and the biggest errors in the estimation of $a_\mu$ arise from the hadronic vacuum polarization~\cite{Jegerlehner:2017gek,Davier:2017zfy,Keshavarzi:2018mgv}. 
%The problem which we are confronted with is two-folded.
One of the  main obstacles to reduce the error  of the hadronic contribution to $a_\mu$ is the discrepancy between the experimental extractions  of the pion form factor from the cross section of the reaction
$e^+e^-\to \pi^+\pi^-\gamma$ by using the initial state radiation (ISR) method~\cite{Lees:2012cj,Aubert:2009ad,
Anastasi:2017eio,Babusci:2012rp,Ambrosino:2010bv,Ambrosino:2008aa,
Ablikim:2015orh}.
The most relevant hadronic contribution (about 70\%) to the determination of $a_\mu$ comes from the region
of the $\pi^+\pi^-$ invariant mass around the $\rho-\omega$ resonances. %It amounts to about 70\% of the hadronic contribution to $a_\mu$ in the $\rho-\omega$ region. 
The biggest difference~\cite{Anastasi:2017eio},
between KLOE and BABAR measurements, amounts there to about 2\%. It goes even up to 10\% around the
$\omega$ resonance region, though that region is very narrow and its contribution to $a_\mu$ is smaller. 
For higher  $\pi^+\pi^-$ invariant masses (at 0.9 GeV) the difference raises to $5\%$. 
%\newline

A possible source of the discrepancy might be attributed
to missing radiative corrections in the event generator(s) used
in the experimental analyses since radiative corrections could be different at different
energies of the experiments. The Monte Carlo event generator \texttt{PHOKHARA}~\cite{Rodrigo:2001kf}
was used by all the experiments, both for
the mode $e^+e^-\to \mu^+\mu^-\gamma$, which serves as a luminosity
monitoring process, and the mode $e^+e^-\to \pi^+\pi^-\gamma$,
which was used to extract the cross section of the reaction  $e^+e^-\to \pi^+\pi^-$,  and the corresponding pion form factor.
In Ref.~\cite{Campanario:2013uea}, the complete QED next-to-leading order (NLO) radiative corrections
to the cross section of the reaction  $e^+e^-\to \mu^+\mu^-\gamma$ were calculated and implemented in \texttt{PHOKHARA}.
It was shown there that the radiative corrections for that process, which were missing in the
\texttt{PHOKHARA} event generator at the time of the experimental
analyses, were at least one order of magnitude smaller than the discrepancies
between experiments.
The still missing ISR NNLO corrections were estimated in Ref.~\cite{Rodrigo:2001kf}
to be at most 0.3\%. That estimate was later confirmed in Ref.~\cite{Jadach:2005gx}.
It is taken as a part of the intrinsic accuracy of the \texttt{PHOKHARA} event generator, which is 0.5\%, and is added
in the experimental analysis as a part of the systematic error. To improve and control errors in a better way, in what follows, we implement neglected so far corrections in the \texttt{PHOKHARA} event generator and discuss their impact on 
the determination of the pion form factor for the
realistic experimental cuts used by BABAR, BES-III and KLOE.

\begin{figure}[h!]
\begin{center}
\hskip-.3cm
\includegraphics[width=8.9cm]{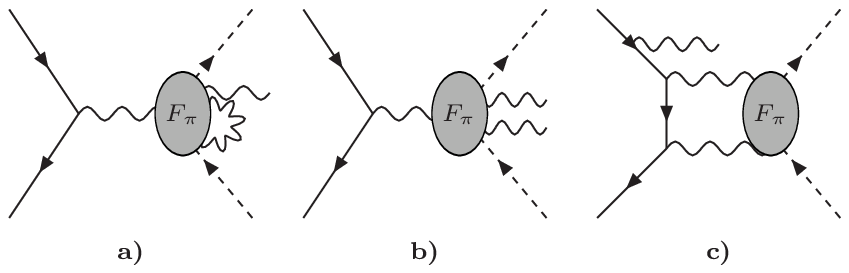}
%\phantom{}\hskip-1cm\includegraphics[width=9.cm]{KLOE_NEW.ps}
\caption{Representative final-state radiation~(\texttt{FSRNLO}) and 
two-virtual-photon~(\texttt{TVP}) diagrams discussed in the text, which describe the radiative corrections to the $e^+e^-\to \pi^+\pi^-\gamma$ process. In diagram a) one photon is radiated from one virtual internal line or external line, in diagram b) two photons are radiated from the final-state pions, in c) two virtual photons couple both to the final-state pions and initial leptons, and one external photon is emitted either from the initial or the final state.
\label{plot0}
}
\end{center}
\end{figure}

\paragraph{Setting.}
%In this letter, we report  on the radiative corrections to the cross section of the reaction $e^+e^-\to \pi^+\pi^-\gamma$ that
%were neglected so far and could have affected the experimental measurements:
In Fig.~\ref{plot0}, the two new classes of radiative corrections to the $e^+e^-\to \pi^+\pi^-\gamma$ cross section discussed in this work are shown, namely, the final-state radiation corrections (\texttt{FSRNLO}) in diagrams a) and b), and the two-virtual-photon (\texttt{TVP}) contributions in c). 
These two types of contributions are separately gauge independent, allowing to show separate results for them.
%that were neglected so far and could have affected the experimental measurements are shown.

The \texttt{TVP} contributions appear for the first time at NLO and 
consist of Feynman diagrams with two virtual photons exchanged between the initial-state electron-positron line and the final-state pions, and one extra real photon emitted either from the initial or the final state. They include up to pentagon topologies.
At the NLO level, only the interference of these corrections with the  Born diagrams contribute to the cross section. 
These corrections are ultraviolet finite and infrared divergent.
%The \texttt{FORTRAN} code, which calculates these corrections, 
%was  generated by a dedicated Mathematica code based on \cite{Campanario:2011cs}.
%The trace method was used to sum over fermion polarizations. 
To control the numerical accuracy, which is critical in some kinematical regions, 
the tensor integrals were calculated using the method described in Refs.~\cite{Denner:2005nn,Binoth:2005ff} for the 5-point functions with the conventions defined in Ref.~\cite{Campanario:2011cs}. The scalar one-loop integrals were calculated with the QCDLoop library~\cite{Ellis:2007qk} and cross checked against the LoopTools library~\cite{Hahn:1998yk} with quadruple precision.
The infrared divergences were regularized in dimensional regularization~\cite{tHooft:1972tcz},
and were cancelled by the appropriate soft photon contributions.

In order to speed up the Monte Carlo event generation, the distributed code works mostly in double precision, and quadruple precision is used only in tensor integrals of \texttt{TVP} contributions  to assure the numerical stability.
%
%     The distributed code calculates the one-loop scalar integrals in double
%     numerical precision. Other parts of the code have to run in quadruple
%     precision to assure numerical stability of the code in the whole allowed
 %    phase space. 
The numerical accuracy of the distributed event generator was also checked 
against an independent code that was generated with
FeynArts~\cite{Hahn:2000kx} and FeynCalc~\cite{Mertig:1990an}.      
Both the scalar and the tensor integrals were calculated
there in quadruple precision by using the LoopTools library~\cite{Hahn:1998yk}.
A perfect agreement between the two codes was found in phase space
points far from the collinear regions. In the collinear regions, where a real
photon is emitted along  the direction of an initial or final state particle, the numerical
accuracy of the distributed code assures 5 significant digits of the result.
Gauge independence tests were also performed, as well as tests checking the
independence of the result on the slicing parameter separating the phase space
of the photon emission into the soft part, where the integral is calculated
analytically, and the hard part, where the integral is performed numerically.
The gauge independence of the matrix elements holds at the level of
$10^{-12}$ relative to the result, while the dependence on the separation parameter
yields a numerical precision of 0.02\%. 

Similar tests were performed for the \texttt{\texttt{FSRNLO}} contributions, 
which consist of one-loop corrections to the final
$\pi^+\pi^-\gamma$ state and the appropriate two photon real emission.
These corrections are both ultraviolet and infrared divergent. 
To cancel the ultraviolet singularities, the renormalization on-shell mass scheme was used.
This part of the code  is more stable numerically and was kept completely in double precision.

%
%
%\section{2. The size of the radiative corrections for experimental event selections}
% \label{size}
\begin{figure}
\begin{center}
\hskip-1cm\includegraphics[width=9.cm]{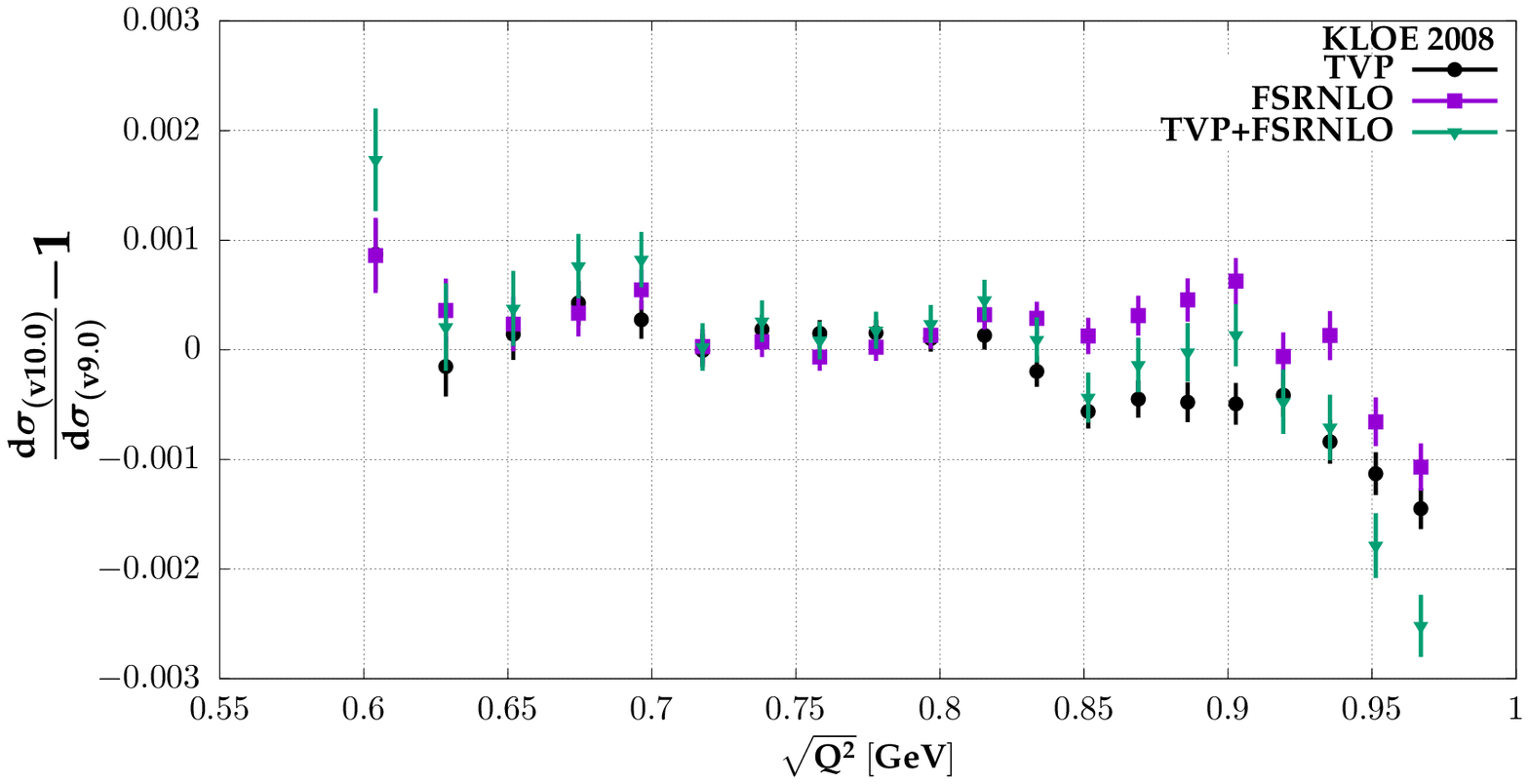}
\phantom{}\hskip-1cm\includegraphics[width=9.cm]{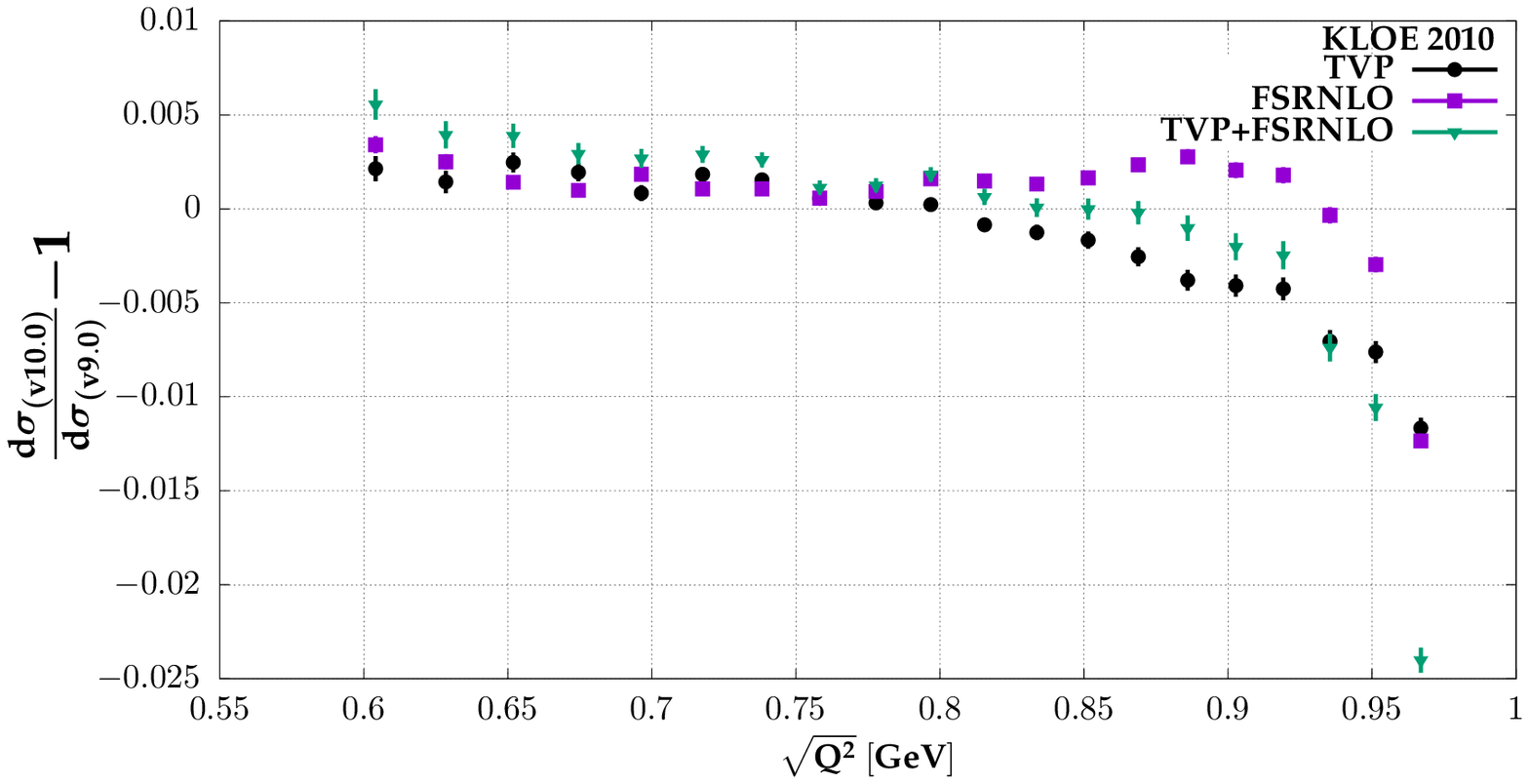}
\caption{The size of the \texttt{TVP} and \texttt{\texttt{FSRNLO}} radiative corrections for KLOE~\cite{Ambrosino:2008aa} (KLOE 2008; photon was not tagged)
  and ~\cite{Ambrosino:2010bv} (KLOE 2010; tagged photon) event selections  as a function of the $\pi^+\pi^-$ invariant mass $Q^2$.
\label{plot1}
}
\end{center}
\end{figure}
 
   \begin{figure}
\begin{center}
\hskip-1cm\includegraphics[width=9.cm]{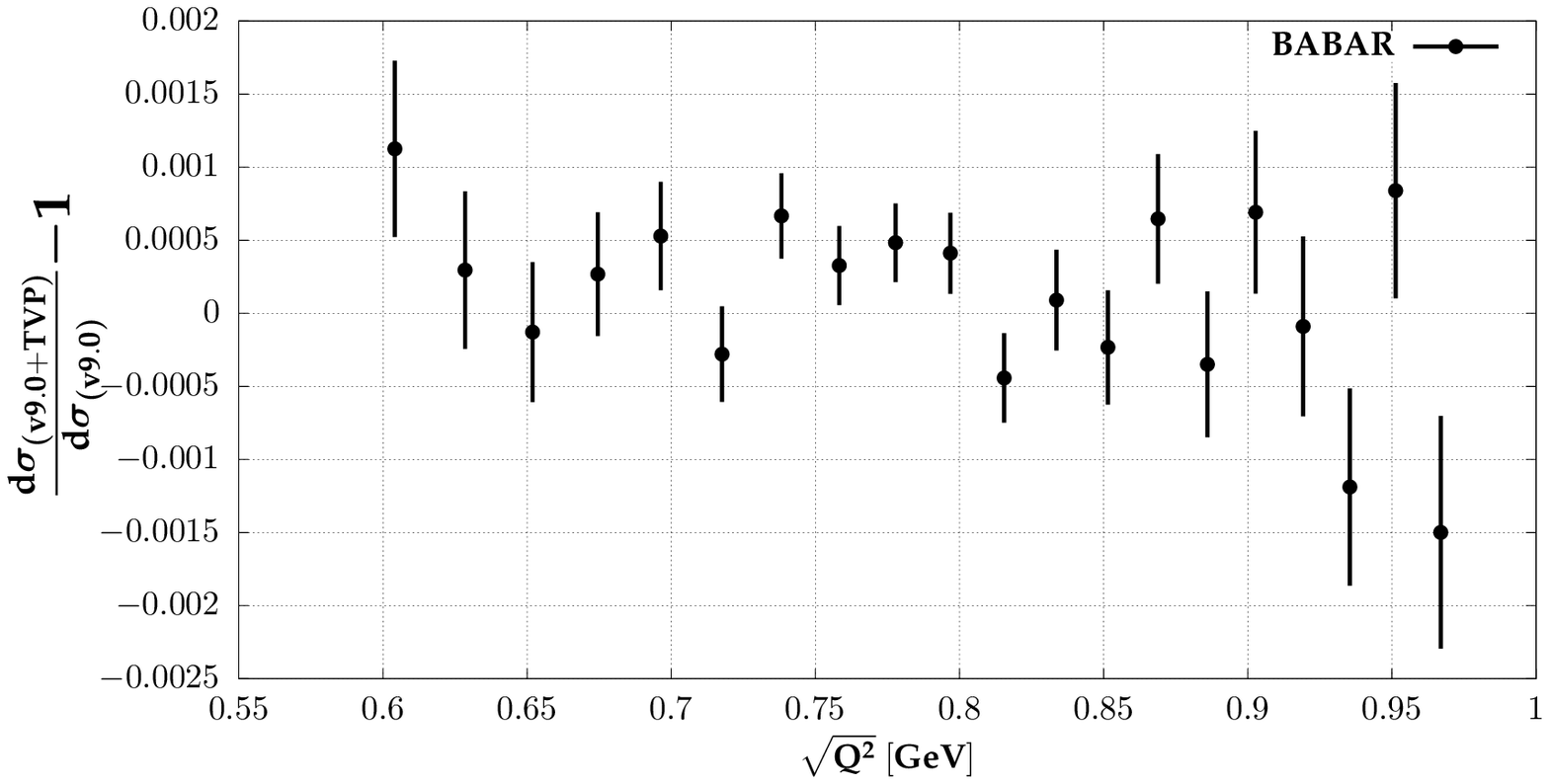}
\phantom{}\hskip-1cm\includegraphics[width=9.cm]{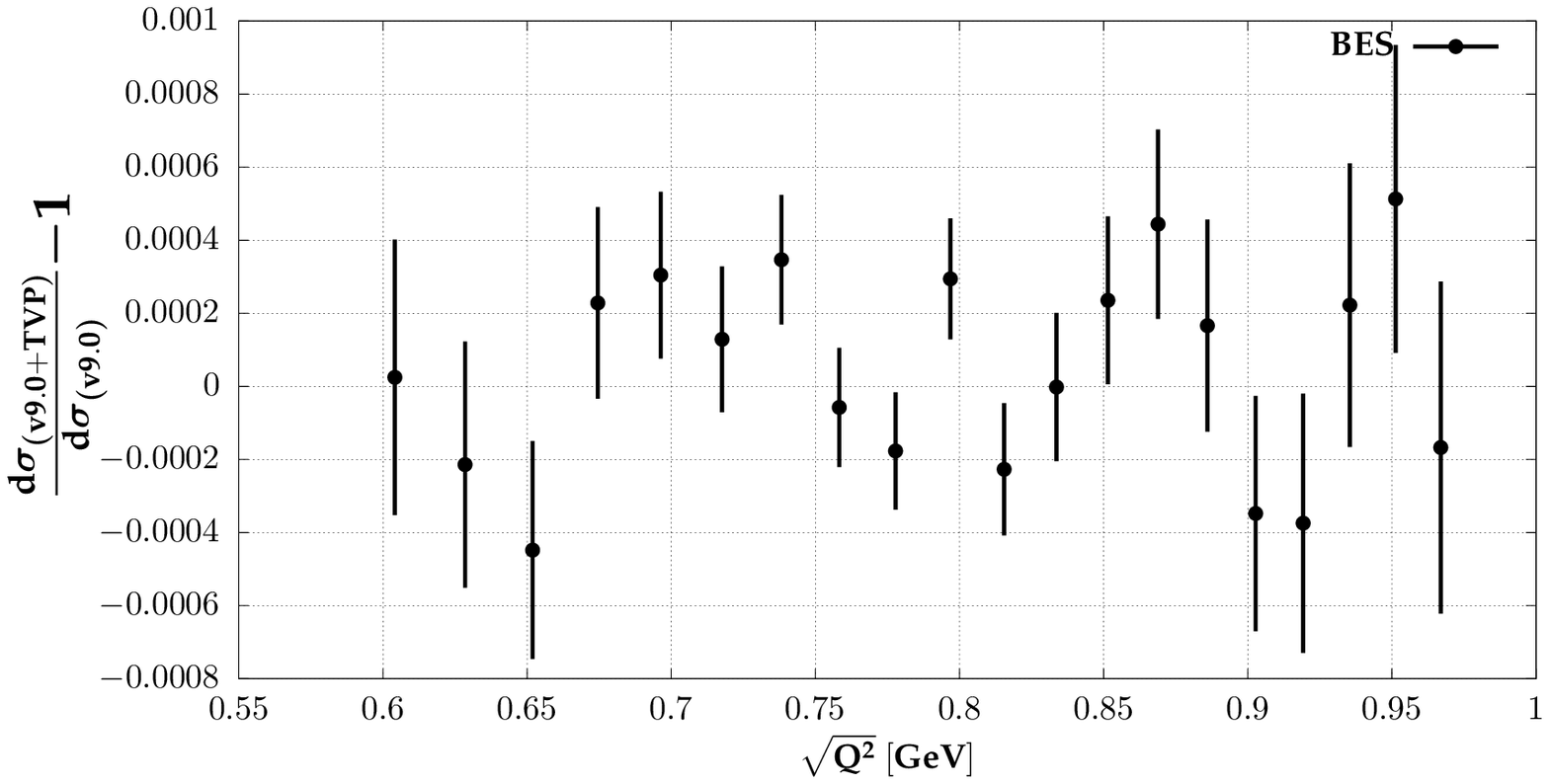}
\caption{The size of the {\texttt{TVP}} radiative corrections for BABAR~\cite{Aubert:2009ad} and BES-III~\cite{Ablikim:2015orh} event selections as a function of the $\pi^+\pi^-$ invariant mass $Q^2$. The \texttt{FSRNLO} contribution is well below $10^{-4}$
   and thus \texttt{TVP}+\texttt{FSRNLO} contribution is almost identical with the \texttt{TVP} contribution.
\label{plot2}
}
\end{center}
   \end{figure}

The  new open source version 10.0 of the \texttt{PHOKHARA} event generator with complete NLO radiative corrections to the cross section of the reaction $e^+e^-\to \pi^+\pi^-\gamma$
is distributed from the web page \cite{wwwphokhara}.
 
\paragraph{Results and discussion.}  
Having the complete radiative corrections implemented into the event
generator \texttt{PHOKHARA}, we can examine how big are these corrections for
the event selections used by the BABAR~\cite{Lees:2012cj,Aubert:2009ad},  KLOE~\cite{Anastasi:2017eio,Babusci:2012rp,Ambrosino:2010bv,Ambrosino:2008aa} 
and BES-III~\cite{Ablikim:2015orh} collaborations.
Their contribution is compared to the predictions of \texttt{PHOKHARA} v9.0~\cite{Campanario:2013uea}, which was used in the experimental analysis of Ref.~\cite{Ablikim:2015orh}.
In all other analysis~\cite{Lees:2012cj,Aubert:2009ad,Anastasi:2017eio,Ambrosino:2008aa,Ambrosino:2010bv,Babusci:2012rp}, earlier versions of \texttt{PHOKHARA} were used, which 
provide identical results as v9.0 for the channel $e^+e^-\to \pi^+\pi^-\gamma$.
The most relevant region for the evaluation of the muon anomalous magnetic moment is the pion pair invariant mass $Q^2$ in the range 0.6-0.9 GeV.
This range was also used in the comparison between experiments in the latest KLOE compilation \cite{Anastasi:2017eio}. We will separately discuss the results within and outside this relevant range ({RR}).

For KLOE, the relative impact of the \texttt{TVP} and \texttt{FSRNLO}
radiative corrections, as implemented in \texttt{PHOKHARA} v10.0,
is shown in Fig.~\ref{plot1}. 
For the KLOE 2008 event selection~\cite{Ambrosino:2008aa}, 
they are both below 0.10\% in the RR, and their sum 
amounts up to 0.18\% at low pion pair invariant masses. At the $\rho$
peak, they are well below 0.05\%. Similar results are expected for the event
selections used by KLOE in Ref.~\cite{Babusci:2012rp}
as the event selection is almost identical to
KLOE 2008~\cite{Ambrosino:2008aa}. For the KLOE 2010 measurement
with a photon tagged in the detector~\cite{Ambrosino:2010bv}, the
radiative corrections can be larger, up to 0.5\% in the RR, for both the \texttt{TVP}
and the \texttt{FSRNLO} contributions. The sum is also at most 0.5\%. Again around
the $\rho$ peak they are smaller and amount up to 0.2\%.
Above 0.9 GeV, which is outside the RR, the corrections can be much bigger
reaching up to 2.4\%. It shows that if one aims to improve the accuracy
in this region a dedicated study of the FSR by both experimental and theory
groups is necessary.

For  the  BABAR~\cite{Lees:2012cj,Aubert:2009ad} and BES-III~\cite{Ablikim:2015orh}
event selections, the size of the {\texttt{TVP}} corrections are shown in Fig.~\ref{plot2}.
They are below $0.10\%$ in both cases. The \texttt{FSRNLO} corrections at the energies of these
experiments are at least two orders of magnitude smaller than for KLOE, and are thus negligible.
The reason for this is that they are proportional to the modulus
square of the pion form factor evaluated at the energy of the given experiment
and the form factor falls rapidly with the energy.
 
The radiative corrections involving pions are intrinsically model dependent.
Yet, even if we
conservatively assume that the discrepancy is about 50\% of the obtained result,
they cannot by any means explain the above mentioned differences between
the experimental measurements.
%Actually, such a safe assumption has been applied in the $a_\mu$ estimate. As we can see,
The actual accuracy of the presented
results is much better than the 50\% mentioned above,
as the model used here was well tested experimentally leaving no space for substantial deviations (see Ref.~\cite{Actis:2010gg} for discussion and further references). A good agreement  with the data was found, while additional dedicated tests 
would be required if a more accurate estimate of the model dependence is needed.
{This is especially important for KLOE 2010 with the pion pair invariant mass range above 0.9 GeV.}

\paragraph{Summary and Outlook.}
We
conclude that the last set of NLO radiative corrections not considered earlier in the event generator \texttt{PHOKHARA},
which was used by the BABAR, KLOE and BES-III collaborations, cannot be  the source of
the discrepancies between the different extractions of the pion form factor performed by these
experiments.  As a consequence, these corrections cannot be the origin of the discrepancy between the 
experimental measurement and the SM prediction of the muon anomalous magnetic moment $a_\mu$ because they are too small. More effort is needed on the experimental side and further, more accurate measurements of the pion form factor are needed to resolve that long standing puzzle, 
and also to match the expected precision attainable at the next generation of $a_\mu$ experiments.
With this work, a new version of the event generator \texttt{PHOKHARA} with complete NLO radiative corrections is available for more refined future measurements of the pion form factor.

\paragraph{Acknowledgements-}
The work supported in part by
the Polish National Science Centre, grant number DEC-2012/07/B/ST2/03867,
the German Research Foundation DFG under Contract No.~Collaborative Research Center CRC-1044, 
the Generalitat Valenciana, Spanish Government and ERDF funds from the European Commission 
(Grants No.~PROMETEO/2017/053, FPA2017-84445-P, FPA2017-84543-P  and SEV-2014-0398),
and the COST Action PARTICLEFACE (CA16201).
{\it F.C.}~acknowledges financial support by the Spanish Government and 
Generalitat Valenciana (Grants No.~RYC-2014-16061 and SEJI-2017/2017/019). {\it J.G.} is supported in part by the Polish National Science Centre, grant number DEC-2013/11/B/ST2/04023 and by international mobilities for research activities of the University of Hradec Kr\'alov\'e, CZ.02.2.69/0.0/0.0/16\_027/0008487.

\bibliographystyle{elsarticle-num}
\bibliography{biblio}

\end{document}